\documentclass[aps,pra,twocolumn,twoside,a4paper,showpacs]{revtex4}

\usepackage[tbtags]{amsmath}
\usepackage{bm,graphicx,enumerate,mathptmx}

\DeclareMathOperator{\Tr}{Tr} 
\DeclareMathOperator{\sign}{sgn}
\DeclareSymbolFont{symbols}{OMS}{cmsy}{m}{n}
\renewcommand{\vec}[1]{\mathbf{#1}}

\begin{document}
\bibliographystyle{apsrev}

\title{Einstein-Podolsky-Rosen correlations of spin measurements in
  two moving inertial frames}

\author{Jakub Rembieli{\'n}ski}%
\email{J.Rembielinski@merlin.fic.uni.lodz.pl}%
\affiliation{Katedra Fizyki Teoretycznej, Uniwersytet {\L}{\'o}dzki,
  Pomorska 149/153, 90-236 {\L}{\'o}d{\'z}, Poland}%

\author{Kordian Andrzej Smoli{\'n}ski}%
\email{K.A.Smolinski@merlin.fic.uni.lodz.pl}%
\affiliation{Katedra Fizyki Teoretycznej, Uniwersytet {\L}{\'o}dzki,
  Pomorska 149/153, 90-236 {\L}{\'o}d{\'z}, Poland}%

\date{26 November 2002}

\begin{abstract}
  The formula for the correlation function of spin measurements of two
  particles in two moving inertial frames is derived within
  Lorentz-covariant quantum mechanics formulated in the absolute
  synchronization framework.  These results are the first exact
  Einstein-Podolsky-Rosen correlation functions obtained for
  Lorentz-covariant quantum-mechanical systems in moving frames under
  physically acceptable conditions, i.e., taking into account the
  localization of the particles during the detection and using the
  spin operator with proper transformation properties under the action
  of the Lorentz group.  Some special cases and approximations of the
  calculated correlation function are given.  The resulting
  correlation function can be used as a basis for a proposal of a
  decisive experiment for a possible existence of a quantum-mechanical
  preferred frame.
\end{abstract}
\maketitle

\section{Introduction}
\label{sec:introduction}

Contemporary considerations of the Einstein-Podolsky-Rosen (EPR)
\cite{einstein35,bohm51} correlations are restricted mostly to
observers staying in a fixed inertial frame of reference (for the
theoretical prescriptions and experimental results see, e.g.,
Refs.~\cite{aspect81,aspect82,aspect82b,bernstein93,bouwmeester99,
pan01,stefanov02,tittel98,tittel99,weihs98,zbinden01b}).  %
This is motivated not only by the experimental requirement but, first
of all, because of very serious difficulties connected with
description of EPR-like experiments in frames in a relative motion.
There are two reasons of the troubles with understanding and
calculating the EPR correlation function in this case.  The first one
is related to the relativity of the notion of simultaneity for moving
observers versus instantaneous state reduction.  The second problem is
related to the nonexistence of a covariant notion of localization in
the relativistic quantum mechanics \cite{bacry88}.  The latter
deficiency is especially serious because every realistic measurement
involves localization in the detector area.

Proposed solutions to these problems strongly depend on the adopted
interpretation of quantum mechanics.  From an orthodox point of view,
attribution of physical meaning to the final probabilities only does
not lead to a serious tension between quantum mechanics (QM) and
special relativity, so they can ``peacefully coexist''
\cite{aharonov81,peres95,peres00a,peres00b}.

The second line of understanding of QM lies in attributing a physical
meaning to the physical state, its time evolution, localization, etc.
From this point of view there are serious problems on the border
between quantum mechanics and special relativity
\cite{bell81,hardy92,stapp77,suarez00,scarani02}.  The most important
ones are: lack of the manifest Lorentz covariance of quantum mechanics
with finite degrees of freedom and the above mentioned nonexistence
of a covariant notion of localization.  Troubles with a sharp
localization in the relativistic QM arise also if we restrict
ourselves to a fixed inertial frame (cf.\ Hegerfeld
theorem~\cite{hegerfeld98,hegerfeld01}).

Following Bell \cite{bell81}, a consistent formulation of quantum
mechanics requires a preferred frame (PF) at the fundamental level (it
is interesting that also Einstein and Dirac had admitted such a
``nonmechanical'' notion of a preferred frame
\cite{einstein22,dirac51}).  Bell gives the very clear point of view
on to this question in Ref.~\cite{davies86}.

A conceptual difficulty related to the notion of the PF lies in an
apparent contradiction with the Lorentz symmetry.  In
Refs.~\cite{caban99,rembielinski80,rembielinski97} it was shown that
this is not the case: it is possible to arrange Lorentz group
transformations in such a way that the Lorentz covariance survives
while the relativity principle (democracy between inertial frames) is
broken on the quantum level.  Moreover, such an approach is consistent
with all the classical phenomena.  The physical meaning of the new
form of the Lorentz group transformations lies in new, absolute
synchronization scheme for clocks different from Einstein's scheme
\cite{rembielinski80,rembielinski97,reichenbach69,jammer79,will92,will93,%
mansouri77,anderson98}.  %
Both synchronizations, the new and the standard one, are physically
inequivalent on the classical level only for velocities greater than
the velocity of light.  Furthermore, the causality notion, which is
implied by the nonstandard absolute synchronization, is more general
than the Einstein one and thus it is applicable to nonlocal phenomena.
A Lorentz-covariant formulation of QM based on the above mentioned
absolute synchronization scheme was given in \cite{caban99}.  In such
a formalism it is possible to define the Lorentz covariant notion of
localization and spin, i.e., covariant localized states and a
covariant position operator as well as the spin operator transforming
properly under the action of the Lorentz group.  Note, that exactly
these notions are relevant to a correct discussion of (non) locality
in QM.

A serious candidate for a PF is the cosmic background radiation frame
(CBRF); this choice is connected with possible dynamical
(cosmological) distinguishing of a local privileged frame.  Most
recent EPR experiments performed in Geneva \cite{scarani00} have been
analyzed according to PF hypothesis \cite{caban99} and give a lower
bound for the speed of ``quantum information'' in CBRF at $2 \times
10^{4}c$.  Moreover some attention was also devoted to PF as a
consequence of a possible breaking of the Lorentz invariance
\cite{coleman99,colladay98} in high-energy processes.

Since the covariant spin operator also exists in the formulation of QM
based on the absolute synchronization scheme \cite{caban99}, therefore
we can calculate precisely the EPR correlation function for any spin.
To our knowledge, our results are the first exact EPR correlation
functions obtained for Lorentz-covariant quantum-mechanical systems in
moving frames under physically acceptable conditions (some attempts
were given in interesting papers by Czachor
\cite{czachor97a,czachor97b}; see however Ref.~\footnote{In
  Refs.~\cite{czachor97a,czachor97b} an average value of the operator
  $a_\mu W^\mu \otimes b_\nu W^\nu$, where $W^\mu$ is Pauli--Lubanski four-vector, was
  calculated.  However, in our opinion, this average cannot be treated
  as a correlation function of spin measurements in the relativistic
  case because the spatial part of $W^\mu$ is not the spin operator in
  QM ($\vec{W}$ belongs to the enveloping algebra of Lie algebra of
  Poincar{\'e} group, while the spin operator is a generator of the
  intrinsic rotations---see, e.g., Ref.~\cite{bogolubov75}).
  Moreover, the derivation of correlation function in
  Refs.~\cite{czachor97a,czachor97b} does not involve localization of
  measured particles in detectors and is restricted to the
  measurements performed in the same inertial frames.}).  Because the
resulting formula for the correlation function depends on the
velocities of the preferred frame it can also help us to answer the
old question concerning the existence of a PF by means of the quantum
mechanical EPR experiment and possibly solve the dilemma posed by
Bell.

\section{Preliminaries}
\label{sec:pereliminaries}

\subsection{Realizations of the Lorentz group in the absolute 
  synchronization scheme}
\label{sec:real-lorentz-group}

In this section we briefly describe main features of the absolute
synchronization scheme mentioned above which is used in this work.
The derivation of the presented results can be found in
Refs.~\cite{caban99,rembielinski80,rembielinski97}.  The main idea is
based on a well-known fact that the definition of time coordinate
depends on the procedure used to synchronize clocks \cite{jammer79}.
If we restrict ourselves to the timelike or lightlike signal
propagation, the choice of this procedure is a convention
\cite{jammer79,mansouri77,reichenbach69,will92,will93,anderson98}.
Now, the form of Lorentz transformations depends on the
synchronization scheme, and we can find a synchronization procedure
which leads to the desired form of Lorentz transformation preserving
instant time (i.e., $x^0 = \mathrm{const}$) hyperplanes.  To perform
such a program one has to distinguish an inertial frame, called the
preferred frame: Every absolute synchronization scheme (ASS)
distinguishes formally such a priviledged inertial frame.  We can go
from one ASS to another by the action of the so-called synchronization
group \cite{caban99,rembielinski97}.  The classical relativity
principle can be formulated in this language as the invariance of
physical laws under the action of the synchronization group, or more
simply, by the statement that each inertial frame can be chosen as the
preferred frame, i.e., the choice of the preferred frame is physically
irrelevant.  The very serious advantage of ASS is the separation of
the two fundamental notions of special relativity, namely, the
relativity principle and the Lorentz covariance.  In the absolute
synchronization scheme, even in the case if the relativity principle
is broken, the Lorentz symmetry survives.

Now, each inertial frame is determined by its four-velocity with
respect to the preferred one.  We shall denote the four-velocity of
the preferred frame as seen by an observer at rest in an inertial
frame by $u = (u^0, \vec{u})$.

According to Refs.~\cite{rembielinski80,rembielinski97} the
transformation of the coordinates between inertial frames $O_u$ and
$O_{u'}$ takes the following form:
\begin{subequations}
  \label{eq:1}
  \begin{equation}
    \label{eq:1a}
    x'(u') = D(\Lambda, u) x(u),
  \end{equation}
  where $\Lambda$ is an element of the Lorentz group, $u$ is the
  four-velocity of the preferred frame with respect to $O_u$, and
  $D(\Lambda, u)$ is a $4 \times 4$ matrix depending on $\Lambda$ and $u$.  This
  equation must be accompanied by the transformation law for the
  four-velocity of a preferred frame, which [according to
  Eq.~\eqref{eq:1a}] takes the form
  \begin{equation}
    \label{eq:1b}
    u' = D(\Lambda, u) u.
  \end{equation}
\end{subequations}
We point out that both Eqs.~\eqref{eq:1} are written for contravariant
components of coordinate and four-velocity.

The explicit form of the matrix $D(\Lambda, u)$ is (see
Refs.~\cite{caban99,rembielinski97}), for rotations
\begin{subequations}
  \label{eq:2}
  \begin{equation}
    \label{eq:2a}
    D(R, u) = \left(
      \begin{array}{c|c}
        1&0\\
        \hline
        0&R
      \end{array}
    \right),
  \end{equation}
  where $R \in \mathrm{SO}(3)$ is a standard rotation matrix, and for boosts,
  \begin{equation}
    \label{eq:2b}
    D(w, u) = \left(
      \begin{array}{c|c}
        (w^0)^{-1}&0\\
        \hline
        -\vec{w}&
        I + \dfrac{\vec{w} \otimes \vec{w}^T}{1 + \sqrt{1 + |\vec{w}|^2}} 
        - u^0 \vec{w} \otimes \vec{u}^T
      \end{array}
    \right),
  \end{equation}
\end{subequations}
where $w = (w^0, \vec{w})$ denotes a four-velocity of the frame
$O_{u'}$ as seen by the observer in the frame $O_u$.

Hereafter we use the natural system of units with $c = \hbar = 1$.

Transformations \eqref{eq:1} leave the line element $ds^2 = g_{\mu\nu}(u)
dx^\mu\, dx^\nu$ invariant, where
\begin{displaymath}
  [g_{\mu\nu}(u)] = \left(
    \begin{array}{c|c}
      1& u^0 \vec{u}^T\\
      \hline
      u^0 \vec{u}& -I + (u^0)^2 \vec{u} \otimes \vec{u}^T
    \end{array}
  \right).
\end{displaymath}
Notice that $g(u)$ is constant (i.e., $x$ independent) in each
inertial frame and is congruent to the Minkowskian metric $\eta =
\mathrm{diag}(+,-,-,-)$.  It is easy to check that the space element
is Euclidean, i.e., $dl^2 = d\vec{x}^2$.

The four-velocities $u$, $u'$, and $w$ are related by
\begin{equation}
  \label{eq:3}
  w^0 = \frac{u^0}{{u'}^0}, \quad 
  \vec{w} = \frac{(u^0 + {u'}^0) (\vec{u} - \vec{u}')}{1 
    + u^0 {u'}^0 (1 + \vec{u} \cdot \vec{u}')}.
\end{equation}

The relation between coordinates in the standard and the absolute
synchronization is given by
\begin{subequations}
  \label{eq:4}
  \begin{align}
    x^0_E &= x^0 + u^0 \vec{u} \cdot \vec{x}, & \vec{x}_E &= \vec{x},\\
    u^0_E &= (u^0)^{-1}, & \vec{u}_E &= \vec{u},
  \end{align}
\end{subequations}
where the subscript $E$ indicates coordinates in the standard
(Einstein's) synchronization, while the coordinates in the absolute
synchronization are written without any subscript.  We see that only
the time coordinate changes.  Note also that in the same point of
space we have $\Delta x^0_E = \Delta x^0$, so the time lapse is the same in both
synchronizations.  The coordinates $x_E$ in the Einstein's
synchronization transform according to the standard law, i.e.,
${x'}^\mu_E = \Lambda^\mu{}_\nu x^\nu_E$.

It is important to stress that the transformations \eqref{eq:1} form a
realization of the Lorentz group which transforms linearly space-time
coordinates according to Eq.~\eqref{eq:1a} and simultaneously,
nonlinearly transforms the PF four-velocity according to
\eqref{eq:1b}.  The round-trip velocity of light is invariant under
Eqs.~\eqref{eq:1}.  In particular, the Reichenbach synchronization
coefficient \cite{reichenbach69,jammer79} is given by $\epsilon(\vec{n},
\vec{u}) = (1 - u^0 \vec{n} \cdot \vec{u})/2$.  Moreover, from
Eqs.~\eqref{eq:4} we have the following relation between velocities in
the absolute and the standard synchronizations:
\begin{subequations}
  \label{eq:5}
  \begin{equation}
    \label{eq:5a}
    \vec{v} = \frac{\vec{v}_E}{1 - \dfrac{\vec{v}_E \cdot \vec{u}_E}{u^0_E}},
  \end{equation}
  \begin{equation}
  \label{eq:5b}
    \vec{v}_E = \frac{\vec{v}}{1 + u^0 \vec{v} \cdot \vec{u}}.
  \end{equation}
\end{subequations}
Notice that for $|\vec{v}_E| > 1$ the above formulas have
singularities, i.e., if a superluminal propagation (possibly related
to the nonlocality of the theory) takes place then both descriptions
are no longer equivalent and consequently an ASS is physically
distinguished in such a case even on the classical level
\cite{rembielinski97}.  It is remarkable that the velocity manifold of
spacelike particles is a proper carrier space for the Lorentz group
only in an ASS \cite{rembielinski97}.

We point out that the triangular form \eqref{eq:2b} of a boost matrix
implies that under Lorentz transformations the time coordinate is only
rescaled by a positive factor, i.e., ${x'}^0 = x^0/w^0$, so the time
ordering of events cannot be inverted by any Lorentz transformations
between inertial frames, regardless of the space-time separation.
This is important in the QM context because the transformations of
time do not involve position operators.

\subsection{Lorentz covariant quantum mechanics}
\label{sec:poinc-covar-quant}

The Lorentz-covariant QM was discussed in the framework of an absolute
synchronization scheme in Ref.~\cite{caban99}.  We associate with each
inertial observer in $O_u$ a Hilbert space $\mathcal{H}_u$, so we have
a bundle of Hilbert spaces rather than a single Hilbert space of
states.  It has been shown in Ref.~\cite{caban99} that one can
introduce Hermitian momentum and coordinate four-vector operators
satisfying
\begin{subequations}
  \label{eq:6}
  \begin{align}
    \label{eq:6a}
    [\hat{x}^\mu(u), \hat{p}_\nu(u)] 
    &= i \left(\frac{u_\nu \hat{p}^\mu(u)}{u_\lambda  \hat{p}^\lambda(u)} - \delta^\mu_\nu\right),\\ {}
    \label{eq:6b}
    [\hat{p}_\mu(u), \hat{p}_\nu(u)] &= 0,\\
    \label{eq:6c}
    [\hat{x}^\mu(u), \hat{x}^\nu(u)] &= 0.
  \end{align}
\end{subequations}
We see that $\hat{x}^0$ commutes with all the observables.  This
allows us to interpret $\hat{x}^0$ as a parameter just like in the
standard nonrelativistic quantum mechanics.  Moreover, for
$\Hat{\Vec{x}}$ Eq.~\eqref{eq:6a} is equivalent to $[\hat{x}^i,
\hat{p}_k] = i \delta^i_k$ and $[\hat{x}^i, \hat{p}_0] =
\hat{p}^i/\hat{p}^0$ (notice that the covariant components $u_i = 0$
in each frame), i.e., it has the standard form.  We stress that the
commutation relations \eqref{eq:6} are \emph{covariant} in the
absolute synchronization.  In fact, we have the following
transformation law for four-vector operators
\begin{subequations}
  \label{eq:7}
  \begin{align}
    \label{eq:7a}
    U(\Lambda) \hat{x}^\mu(u) U^{\dag}(\Lambda) &= [D^{-1}(\Lambda, u)]^\mu{}_\nu \hat{x}^\nu(u'),\\
    \label{eq:7b}
    U(\Lambda) \hat{p}_\mu(u) U^{\dag}(\Lambda) &= [D^T(\Lambda, u)]_\mu{}^\nu \hat{p}_\nu(u'),
  \end{align}
\end{subequations}
where $u' = D(\Lambda, u) u$ and $D(\Lambda, u)$ is given by Eqs.~\eqref{eq:2}.
Using Eqs.~\eqref{eq:7} we can transform Eqs.~\eqref{eq:6} to another
reference frame.  We point out once again that under transformations
\eqref{eq:7a} for the time component $\hat{x}^0$ does not mix with
spatial components $\hat{x}^k$ ($k = 1, 2, 3$).  One can also check
that
\begin{equation}
  \label{eq:8}
  [\hat{x}^\mu(u), \hat{p}^2(u)] = [\hat{p}_\mu(u), \hat{p}^2(u)] = 0,
\end{equation}
which means that a localized state has a definite mass.  It is
important to stress that the unitary map which connects one choice of
ASS to another choice of ASS and preserves Eqs.~\eqref{eq:6}
and~\eqref{eq:7} does not exist (this means that the synchronization
group \cite{caban99,rembielinski97} cannot be unitarily realized in
this case).  For this reason QM distinguishes an ASS, i.e., a
particular preferred frame---the quantum preferred frame.  In
Ref.~\cite{rembielinski97} it was shown that the choice of the quantum
preferred frame can be done by the spontaneous breaking of the
synchronization group.  As it was mentioned earlier, a natural
candidate for quantum preferred frame is the CBRF \footnote{Another
  motivation for such a choice is the very recent cosmological
  hypothesis of the so-called rolling tachyon field
  \cite{gibbons02,mukohyama02,sen02}.  In Ref.~\cite{rembielinski97}
  it was proved that quantization of the tachyonic field must be
  necessarily associated with the spontaneous breaking of the
  synchronization group and consequently with the distinguishing of a
  privileged frame.}.

Transformations of the Lorentz group induce an orbit in a bundle of
Hilbert spaces $\mathcal{H}_u$.  Unitary orbits are parameterized by
mass and spin, similarly as for standard unitary representations of
the Poincar{\'e} group.

An orbit induced by an action of the operator $U(\Lambda)$ in the bundle of
Hilbert spaces under consideration is fixed by the following covariant
conditions: (i) $k^2 = m^2$, (ii) $\sign(k^0)$ is invariant; for
physical representations $k^0 > 0$, $\sign(k^0) = 1$.  As a
consequence there exists a positive defined Lorentz-invariant measure
\begin{equation}
  \label{eq:9}
  d\mu(k, m) = d^4k\, \theta(k^0) \delta(k^2 - m^2).
\end{equation}
Now, applying the Wigner method and using Eqs.~\eqref{eq:7} one can
easily determine the action of the operator $U(\Lambda)$ on a basis of
eigenvectors of the four-momentum operator \cite{caban99}
\begin{equation}
  \label{eq:10}
  \hat{p}_\mu(u) |k, u, m; s, \sigma\rangle = k_\mu |k, u, m; s, \sigma\rangle.
\end{equation}
We find \footnote{Here we have chosen slightly different convention
  that in Ref.~\cite{caban99} in the definition of this unitary
  action.}
\begin{equation}
  \label{eq:11}
  U(\Lambda) |k, u, m; s, \sigma\rangle = \mathcal{D}^s(R_{(\Lambda,u)})_{\lambda\sigma} |k', u', m; s, \lambda\rangle,
\end{equation}
where the contravariant components $u^\mu$ and $k^\mu$ transform as
follows
\begin{gather}
  \label{eq:12}
  u' = D(\Lambda, u) u = D(L_{u'}, \tilde{u}) \tilde{u},\\
  \label{eq:13}
  k' = D(\Lambda, u) k,%
\end{gather}
while
\begin{equation}
  \label{eq:14}
  \begin{split}
    R_{(\Lambda,u)} &= D(R_{(\Lambda,u)}, \tilde{u})\\
    &= D^{-1}(L_{u'}, \tilde{u}) D(\Lambda, u) D(L_u, \tilde{u}) \in
    \mathrm{SO}(3).
  \end{split}
\end{equation}
Here $\tilde{u} = (1, \vec{0})$, $u = D(L_u, \tilde{u}) \tilde{u}$ and
$\mathcal{D}^s$ is the standard spin $s$ matrix representation of
$\mathrm{SU}(2)$, $s = 0, \frac{1}{2}, 1,\ldots$; $\sigma, \lambda = -s, -s + 1,\ldots, s -
1, s$.  $R_{(\Lambda,u)}$ is a Wigner rotation belonging to the little group
of a vector $\tilde{u}$.  It should be noted that in this approach,
contrary to the standard one, representations of the Poincar{\'e} group
are induced from the little group of the vector $\tilde{u}$, and not
$\tilde{k} = (m, 0, 0, 0)$.  Finally, the normalization condition for
the basis vectors takes the form
\begin{equation}
  \label{eq:15}
  \langle k, u, m; s, \lambda|k', u, m; s', \lambda'\rangle = 2 k^0 \delta^3(\underline{\vec{k}}' 
  - \underline{\vec{k}}) \delta_{s's} \delta_{\lambda'\lambda},
\end{equation}
where $\underline{\vec{k}}$ denotes the vector formed from covariant
components of the momentum, i.e., $\underline{\vec{k}} = (k_1, k_2,
k_3)$.

\subsection{The localized states and spin}
\label{sec:local-stat-spin}

Following Ref.~\cite{caban99} we construct the localized states (i.e.,
the eigenvectors of the position operator which coincides in PF with
the Newton-Wigner one) and the covariant spin operator.  Eigenstates
of the position operator $\Hat{\Vec{x}}(u)$ (locked up in the $t_0 =
0$) are of the form \cite{caban99}
\begin{equation}
  \label{eq:16}
  |\vec{x}, u, m; s, \sigma\rangle
  = \frac{1}{(2 \pi)^{3/2}} \int \frac{d^3\underline{\vec{k}}}{2
    \omega(\underline{\vec{k}})} \sqrt{u^\lambda k_\lambda}\, e^{i \underline{\vec{k}}
    \cdot \vec{x}} |k, u, m; s, \sigma\rangle,
\end{equation}
where $\omega(\underline{\vec{k}}) = k^0$ is a positive solution of the
dispersion relation $g_{\mu\nu}(u) k^\mu k^\nu = m^2$.  In the Schr{\"o}dinger
picture, after time $t = x^0$ they develop as
\begin{multline}
  \label{eq:17}
  |x^0, \vec{x}, u, m; s, \sigma\rangle\\ 
  = \frac{1}{(2 \pi)^{3/2}} 
  \int \frac{d^3\underline{\vec{k}}}{2 \omega(\underline{\vec{k}})} 
  \sqrt{u^\lambda k_\lambda}\, e^{i k_\mu x^\mu} |k, u, m; s, \sigma\rangle,
\end{multline}
which is not an eigenvector of $\hat{x}(u)$ except for $x^0 = 0$.
These vectors transform under the action of the Lorentz group
according to the following law:
\begin{equation}
  \label{eq:18}
  U(\Lambda) |x^0, \vec{x}, u, m; s, \sigma\rangle = \mathcal{D}^s(R_{(\Lambda,u)})_{\lambda\sigma}
  |{x'}^0, \vec{x}', u', m; s, \lambda\rangle,
\end{equation}
where $x'$ and $u'$ are given by Eqs.~\eqref{eq:1}.  Notice that for
$x^0 = 0$ we have ${x'}^0 = 0$ and ${x'}^k = D(\Lambda, u)^k{}_i x^i$.

Now we define a spin operator \footnote{The operator
  $\hat{\vec{S}}(u)$ is connected with the operator $\hat{S}_{ij}(u)$
  introduced in Ref.~\cite{caban99} by the formula $\hat{S}^i(u) =
  (1/2 u^0) (\delta^{ij} - (u^0)^2/(1 + u^0) u^i u^j) \epsilon^{jkl}
  \hat{S}_{kl}(u).$} in absolute synchronization as follows:
\begin{subequations}
  \label{eq:19}
  \begin{gather}
    \label{eq:19a}
    \Hat{\Vec{S}}(u) |k, u, m; s, \tau\rangle = \bm{\Sigma}^s_{\sigma\tau} |k, u, m; s, \sigma\rangle,%
    \intertext{so}
    \label{eq:19b}
    \Hat{\Vec{S}}(u) |\vec{x}, u, m; s, \tau\rangle 
    = \bm{\Sigma}^s_{\sigma\tau} |\vec{x}, u, m; s, \sigma\rangle,
  \end{gather}
\end{subequations}
where $\bm{\Sigma}^s$ are the standard generators of rotation in the
representation $\mathcal{D}^s$.  The transformation law \eqref{eq:11}
for states implies the following transformation law for the components
of $\Hat{\Vec{S}}(u)$:
\begin{equation}
  \label{eq:20}
  U(\Lambda) \hat{S}^i(u) U^{\dag}(\Lambda) = R^T_{(\Lambda,u)}{}^i{}_j \hat{S}^j(u'),
\end{equation}
where $R_{(\Lambda,u)}$ is a Wigner rotation as above.

Moreover, $\hat{S}^i(u)$ fulfill the standard commutation relations
such that
\begin{equation}
  \label{eq:21}
  [\hat{S}^i(u), \hat{S}^j(u)] = i \epsilon^{ijk} \hat{S}^k(u).
\end{equation}
The invariant $u$-independent spin square operator $\hat{S}^2$ can be
written in terms of $\Hat{\Vec{S}}(u)$ in the standard form
\begin{equation}
  \label{eq:22}
  \hat{S}^2 = (\Hat{\Vec{S}}(u))^2 = s (s + 1) I.
\end{equation}
We stress that only in that formulation of QM it is possible to
introduce the spin operator which transforms properly under the action
of the Lorentz group [see Eq.~\eqref{eq:20}] and satisfies the
standard commutation relations~\eqref{eq:21} \footnote{Recently the
  problem of the spin operator in the standard relativistic QM has
  been discussed in Ref.~\cite{terno}.}.

To analyze the EPR-type experiments we define an observable $\vec{n} \cdot
\Hat{\Vec{S}}(u)$, where
\begin{displaymath}
  \vec{n} = 
  \begin{pmatrix}
    \sin \theta \cos \phi\\
    \sin \theta \sin \phi\\
    \cos \theta
  \end{pmatrix}
  ,
\end{displaymath}
which is the projection of operator $\Hat{\Vec{S}}(u)$ on the
direction of a unit vector $\vec{n}$ in the frame of reference
$\mathcal{O}_u$.  Since $\hat{\vec{S}}$ and $\hat{\vec{x}}$ commute,
i.e., $[\Hat{\Vec{S}}(u), \Hat{\Vec{x}}(u)] = 0$, we can introduce a
set of common eigenvectors of $\Hat{\Vec{x}}(u)$ and $\vec{n} \cdot
\Hat{\Vec{S}}(u)$.  They are given by
\begin{equation}
  \label{eq:23}
  \begin{split}
    |\vec{x}, \vec{n}, u, m; s, \lambda\rangle &= 
    \sqrt{2 u^0} \exp\left(i \theta \vec{e}_{\vec{n}} \cdot \hat{\vec{S}}(u)\right) 
    |\vec{x}, u, m; s, \lambda\rangle\\
    &= \mathcal{D}^s\left(e^{i \theta \vec{e}_{\vec{n}} \cdot \bm{\Sigma}^s}\right)_{\sigma\lambda} 
    \sqrt{2 u^0} |\vec{x}, u, m; s, \sigma\rangle,
  \end{split}
\end{equation}
where
\begin{displaymath}
  \vec{e}_{\vec{n}} = 
  \begin{pmatrix}
    \sin \phi\\
    -\cos \phi\\
    0
  \end{pmatrix}
  .
\end{displaymath}
Vectors \eqref{eq:23} satisfy the following eigenequations:
\begin{subequations}
  \label{eq:24}
  \begin{align}
    \label{eq:24a}
    \Hat{\Vec{x}}(u) |\vec{x}, \vec{n}, u, m; s, \lambda\rangle &= \vec{x}
    |\vec{x}, \vec{n}, u, m; s, \lambda\rangle,\\
    \label{eq:24b}
    \vec{n} \cdot \Hat{\Vec{S}}(u) |\vec{x}, \vec{n}, u, m; s, \lambda\rangle &= \lambda
    |\vec{x}, \vec{n}, u, m; s, \lambda\rangle,
  \end{align}
\end{subequations}
with the normalization
\begin{multline}
  \label{eq:25}
  \langle\vec{x}, \vec{a}, u, m; s, \lambda|\vec{y}, \vec{b}, u, m; s, \sigma\rangle\\ 
  = \delta^3(\vec{x} - \vec{y}) \mathcal{D}^s\left(e^{-i \theta_{\vec{a}} \vec{e}_{\vec{a}} \cdot
      \bm{\Sigma}} e^{i \theta_{\vec{b}} \vec{e}_{\vec{b}} \cdot \bm{\Sigma}}\right)_{\lambda\sigma}.
\end{multline}  
Thus the projector corresponding to a region $\Omega$ and to a value $\lambda$ of
the spin component in the $\vec{n}$ direction in the frame
$\mathcal{O}_u$ is of the form
\begin{equation}
  \label{eq:26}
  P^\lambda_{\Omega,\vec{n}}(u) = \int_\Omega d^3\vec{x}\, |\vec{x}, \vec{n}, u, m; s, \lambda\rangle 
  \langle\vec{x}, \vec{n}_, u, m; s, \lambda|.
\end{equation}
Now, in the Schr{\"o}dinger picture projectors $P^\lambda_{\Omega,\vec{n}}(u)$,
locked up in $x^0 = 0$, are time independent and transform under
Lorentz group transformations by means of Eqs.~\eqref{eq:5}
and~\eqref{eq:23} as follows:
\begin{equation}
  \label{eq:27}
  U(\Lambda) P^\lambda_{\Omega,\vec{n}}(u) U^{\dag}(\Lambda) = P^\lambda_{\Omega',\vec{n}'}(u'),
\end{equation}
here $\vec{n}' = R_{(\Lambda, u)} \vec{n}$ and the region $\Omega'$ is obtained
from the region $\Omega$ by ${x'}^k = D(\Lambda, u)^k{}_i x^i$.  We stress that
there is no analog of Eq.~\eqref{eq:27} in the standard formulation of
relativistic QM.

\section{EPR Correlations}
\label{sec:epr-correlations}

In this section we employ the formalism introduced above to the
calculation of the correlation function of the EPR-type experiment.
We consider distinguishable particles (the case of identical particles
is quite analogous).  In this case vectors describing pure states
belong to $\mathcal{H}^{s_\alpha}_\alpha(u) \otimes \mathcal{H}^{s_\beta}_\beta(u)$, where
indices $\alpha$ and $\beta$ denote particles.  We associate with the observers
$\mathcal{A}$ and $\mathcal{B}$ the two frames $\mathcal{A}_{u_A}$ and
$\mathcal{B}_{u_B}$, the preferred frame four-velocities with respect
to $\mathcal{A}_{u_A}$ and $\mathcal{B}_{u_B}$ are $u_A$ and $u_B$,
respectively.  These observers measure the spin component in the
$\vec{a}$ and $\vec{b}$ directions, respectively ($|\vec{a}| =
|\vec{b}| = 1$), in the space regions $A$ and $B$, respectively.  Let
us denote their observables as $M_{A,\vec{a}}$ and $M_{B,\vec{b}}$,
respectively.  If we assume that the observer $\mathcal{A}$ registers
the particle $\alpha$ and the observer $\mathcal{B}$ registers the particle
$\beta$, then
\begin{subequations}
  \label{eq:28}
  \begin{align}
    \label{eq:28a}
    M_{A,\vec{a}}(u_A) &= \sum_{\mu_\alpha = -s_\alpha}^{s_\alpha} \mu_\alpha P^{\mu_\alpha}_{A,\vec{a}}(u_A) \otimes I
    \equiv \sum_{\mu_\alpha = -s_\alpha}^{s_\alpha} \mu_\alpha \Pi^{\mu_\alpha}_{A,\vec{a}},\\
    \label{eq:28b}
    M_{B,\vec{b}}(u_B) &= I \otimes \sum_{\mu_\beta = -s_\beta}^{s_\beta} \mu_\beta P^{\mu_\beta}_{B,\vec{b}}(u_B)
    \equiv \sum_{\mu_\beta = -s_\beta}^{s_\beta} \mu_\beta \Pi^{\mu_\beta}_{B,\vec{b}},
  \end{align}
\end{subequations}
where $P^{\mu_\alpha}_{A,\vec{a}}$ and $P^{\mu_\beta}_{B,\vec{b}}$ are given by
Eq.~\eqref{eq:26}.

A state of the system under consideration in frame $O_u$ at a time $t$
is denoted by $\rho(u, t)$.  Now we write down the sequence of events
describing the development of the state $\rho(u, t)$.
\begin{enumerate}[(1)]
\item The observer $\mathcal{A}$ performs measurement with selection
  of the spin component $\mu_\alpha$ in the direction $\vec{a}$, localizing
  the particle in the space region $A$ at a time $t^1_A$.  This causes
  the following state reduction
  \begin{displaymath}
    \rho(u_A, t^1_A) \mapsto \frac{\Pi_{A,\vec{a}}^{\mu_a} \rho(u_A, t^1_A) \Pi_{A,\vec{a}}^{\mu_a}}
    {\Tr[\rho(u_A, t^1_A) \Pi_{A,\vec{a}}^{\mu_a}]}
    \equiv \rho_A(u_A, t^1_A; \mu_a).
  \end{displaymath}
\item The observer $\mathcal{B}$ sees the state $\rho_A(u_A, t^1_A; \mu_a)$
  at a time $t^1_B$ as
  \begin{displaymath}
    \rho_A(u_B, t^1_B; \mu_a) = U(\Lambda) \rho_A(u_A, t^1_A; \mu_a) U^{\dag}(\Lambda),
  \end{displaymath}
  where $x_B = D(\Lambda, u_A) x_A$, $u_B = D(\Lambda, u_A) u_A$; so
  $t^1_B = D(\Lambda, u_A)^0{}_0 t^1_A$.
\item The state evolves freely in time from $t^1_B$ to $t^2_B$ to
  \begin{displaymath}
    \rho_A(u_B, t^2_B; \mu_a) = U(t^2_B - t^1_B) \rho_A(u_B, t^1_B; \mu_a) U^{\dag}(t^2_B - t^1_B).
  \end{displaymath}
\item The observer $\mathcal{B}$ performs measurement with selection
  of the observable $M_{B,\vec{b}}$ at a time $t^2_B$
  \begin{displaymath}
    \begin{split}
      \rho_A(u_B, t^2_B; \mu_a) &\mapsto \frac{\Pi_{B,\vec{b}}^{\mu_b} \rho_A(u_B, t^2_B; \mu_a) \Pi_{B,\vec{b}}^{\mu_b}}
      {\Tr[\rho_A(u_B, t^2_B; \mu_a) \Pi_{B,\vec{b}}^{\mu_b}]}\\
      &\equiv \rho_{AB}(u_B, t^2_B; \mu_b|\mu_a).
    \end{split}
  \end{displaymath}
\end{enumerate}

Recall that in the absolute synchronization $t^2_B - t^1_B = D(\Lambda,
u_A)^0{}_0 (t^2_A - t^1_A)$ and $D(\Lambda, u_A)^0{}_0 > 0$, so the causal
relationship between measurements in $\mathcal{A}$ and $\mathcal{B}$
is well established (contrary to the Einstein's synchronization).

Therefore the probability $p(\mu_a)$ that the observer $\mathcal{A}$ has
measured value $\mu_a$ and the probability $p(\mu_b|\mu_a)$ that the
observer $\mathcal{B}$ has measured the value $\mu_b$ if the observer
$\mathcal{A}$ had measured $\mu_a$ are
\begin{widetext}
\begin{displaymath}
  p(\mu_a) = \Tr[\rho(u_A, t^1_A) \Pi_{A,\vec{a}}^{\mu_a}],
\end{displaymath}
\begin{align*}
  p(\mu_b|\mu_a) &= \Tr[\rho_A(u_B, t^2_B; \mu_a) \Pi_{B,\vec{b}}^{\mu_b}] \\
  &= \frac{\Tr\left[\rho(u_A, t^1_A) \Pi_{A,\vec{a}}^{\mu_a} U^{\dag}(\Lambda) U^{\dag}(t^2_B
      - t^1_B) \Pi_{B,\vec{b}}^{\mu_b} U(t^2_B - t^1_B) U(\Lambda)
      \Pi_{A,\vec{a}}^{\mu_a}\right]}{p(\mu_a)},
\end{align*}
thus
\begin{displaymath}
  p(\mu_a) p(\mu_b|\mu_a) = \Tr\left[\rho(u_A, t^1_A) \Pi_{A,\vec{a}}^{\mu_a} U^{\dag}(\Lambda)
    U^{\dag}(t^2_B - t^1_B) \Pi_{B,\vec{b}}^{\mu_b} U(t^2_B - t^1_B) U(\Lambda)
    \Pi_{A,\vec{a}}^{\mu_a}\right].
\end{displaymath}
\end{widetext}
Therefore the correlation function reads
\begin{equation}
  \label{eq:29}
  \begin{split}
    C(\vec{a}, \vec{b}) &= \sum_{\mu_a,\mu_b} \mu_a \mu_b p(\mu_a) p(\mu_b|\mu_a) \\
    &= \sum_{\mu_a} \mu_a \Tr\left[\rho(u_A, t^1_A) \Pi_{A,\vec{a}}^{\mu_a} U^{\dag}(\Lambda)
      U^{\dag}(t^2_B - t^1_B)\right.\\
    &\quad\left. \times M_{B,\vec{b}} U(t^2_B - t^1_B) U(\Lambda)
      \Pi_{A,\vec{a}}^{\mu_a}\right].
  \end{split}
\end{equation}
Recall that in $\mathcal{H}_\alpha \otimes \mathcal{H}_\beta$, $U(\Lambda) = U(\Lambda)_\alpha \otimes
U(\Lambda)_\beta$ and for the free evolution $U(t) = U(t)_\alpha \otimes U(t)_\beta$.

\section{Correlation Function---a Particular Case}
\label{sec:corr-funct-part}

In this section we discuss the case when the measurements in
$\mathcal{A}$ and $\mathcal{B}$ are simultaneous. So we assume that
$t^2_B = t^1_B \equiv t_B$ [i.e., there is no free evolution of a state
between measurements 1 and 2, so $U(t^2_B - t^1_B) = I$].  Moreover,
we assume that the regions $A$ and $B$ are disjoint.  Therefore in
Eq.~\eqref{eq:29} $M_{B,\vec{b}}$ commutes with $U(\Lambda)
\Pi^{\mu_a}_{A,\vec{a}} U^{\dag}(\Lambda)$ and in this case we have ($t^1_A \equiv t_A$)
\begin{equation}
  \label{eq:30}
  C(\vec{a}, \vec{b}) 
  = \Tr\left[\rho(u_A, t_A) M_{A,\vec{a}} U^{\dag}(\Lambda) M_{B,\vec{b}} U(\Lambda)\right].
\end{equation}

Assume that the initial state is a pure state $|\Psi\rangle \in
\mathcal{H}_\alpha(u_A) \otimes \mathcal{H}_\beta(u_B)$, thus {$\rho(u_A, t_A) = |\Psi\rangle
  \langle\Psi|$}, $\langle\Psi|\Psi\rangle = 1$.  Since in this case
\begin{displaymath}
  C(\vec{a}, \vec{b}) = \langle\Psi|M_{A,\vec{a}}(u_A) U^{\dag}(\Lambda) M_{B,\vec{b}}(u_B) U(\Lambda)|\Psi\rangle,
\end{displaymath}
therefore using $U(\Lambda) = U(\Lambda)_\alpha \otimes U(\Lambda)_\beta$, we find
\begin{multline}
  \label{eq:31}
  C(\vec{a}, \vec{b}) = \sum_{\mu_\alpha,\mu_\beta} \mu_\alpha \mu_\beta \langle\Psi|\left(P^{\mu_\alpha}_{A,\vec{a}}(u_A)\right.\\
  \left.\otimes U(\Lambda^{-1}) P^{\mu_\beta}_{B,\vec{b}}(u_B) U^{\dag}(\Lambda^{-1})\right)|\Psi\rangle.
\end{multline}
Hence, taking into account Eq.~\eqref{eq:27} we obtain
\begin{equation}
  \label{eq:32}
  C(\vec{a}, \vec{b}) = \sum_{\mu_\alpha,\mu_\beta} \mu_\alpha \mu_\beta \langle\Psi|P^{\mu_\alpha}_{A,\vec{a}}(u_A) 
  \otimes P^{\mu_\beta}_{B_{\mathcal{A}},\vec{b}'}(u_A)|\Psi\rangle,
\end{equation} 
where $B_{\mathcal{A}}$ is obtained from the region $B$ by
transformation ${x'}^i = D^{-1}(\Lambda, u_A)^i{}_j x^j$ and $\vec{b}' =
R^T_{(\Lambda, u_A)} \vec{b}$.  Now, using the expansion of the vector $|\Psi\rangle$
such that
\begin{multline}
  \label{eq:33}
  |\Psi\rangle = \sum_{\lambda_\alpha,\lambda_\beta} \int d^3\vec{x} \int d^3\vec{y}\, 2 u^0 
  \psi_{\lambda_\alpha\lambda_\beta}(\vec{x}, \vec{y}, u_A)\\ 
  \times |\vec{x}, u_A, m_\alpha; s_\alpha, \lambda_\alpha\rangle 
  \otimes |\vec{y}, u_A, m_\beta; s_\beta, \lambda_\beta\rangle,
\end{multline}
where 
\begin{equation}
  \label{eq:34}
  \langle\Psi|\Psi\rangle = \int d^3\vec{x} \int d^3\vec{y} \Tr\left[\psi^{\dag}(\vec{x}, \vec{y}, u_A) 
    \psi(\vec{x}, \vec{y}, u_A)\right] = 1,
\end{equation}
(hereafter $\psi$ denotes the matrix $\psi = [\psi_{\lambda_\alpha \lambda_\beta}]$) and using
Eqs.~\eqref{eq:30} and \eqref{eq:31} we get, after some calculations,
the following formula:
\begin{multline}
  \label{eq:35}
  C(\vec{a}, \vec{b}) = \int_A d^3\vec{x} \int_{B_{\mathcal{A}}} d^3\vec{y}\,
  \Tr[\psi^{\dag}(\vec{x}, \vec{y}, u_A) \vec{a} \cdot \bm{\Sigma}^{s_\alpha} 
     \psi(\vec{x}, \vec{y}, u_A)\\
   \times (R^T_{(\Lambda, u_A)} \vec{b}) \cdot {\bm{\Sigma}^{s_\beta}}^T].
\end{multline}

Consider now the case of the spin $s_\alpha = s_\beta = 1/2$.  We can write
then $\psi(\vec{x}, \vec{y}, u_A) = (i/\sqrt{2}) \chi(\vec{x}, \vec{y}, u_A)
\sigma^2$ and $\bm{\Sigma} = \frac{1}{2} \bm{\sigma}$, where $\sigma^i$ ($i = 1, 2, 3$)
are the Pauli matrices.  Thus
\begin{displaymath}
  C(\vec{a}, \vec{b}) = -\frac{1}{4} \int_A d^3\vec{x} \int_{B_{\mathcal{A}}} d^3\vec{y}\, 
  |\chi(\vec{x}, \vec{y}, u_A)|^2 (\vec{a} \cdot R^T_{(\Lambda, u_A)} \vec{b}),
\end{displaymath}
i.e., up to a factor
\begin{equation}
  \label{eq:36}
  C(\vec{a}, \vec{b}) \propto \vec{a} \cdot R^T_{(\Lambda, u_A)} \vec{b}.
\end{equation}
If the orientation of axes in the frames $\mathcal{A}_{u_A}$ and
$\mathcal{B}_{u_B}$ is the same, we need to deal with boosts
$\Lambda(\vec{w})$ only and
\begin{equation}
  \label{eq:36a}\tag{\ref{eq:36}a}
  C(\vec{a}, \vec{b}) \propto \vec{a} \cdot R^T_{(\Lambda(\vec{w}), u_A)} \vec{b},
\end{equation}
where $w^\mu$ are components of four-velocity of the frame
$\mathcal{B}_{u_B}$ with respect to $\mathcal{A}_{u_A}$.

\begin{figure}
  \begin{center}
    \includegraphics[width=\columnwidth]{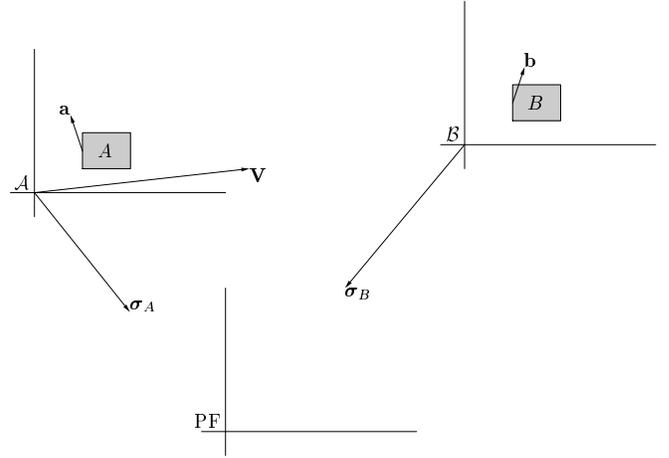}
    \caption{A schematic EPR experiment in moving frames.
      The detectors $A$ and $B$ are at rest in the frames
      $\mathcal{A}$ and $\mathcal{B}$, respectively.  $\bm{\sigma}_A$ and
      $\bm{\sigma}_B$ denote the velocities of PF with respect to
      $\mathcal{A}$ and $\mathcal{B}$, respectively; while $\vec{V}$
      denotes the velocity of $\mathcal{B}$ with respect to
      $\mathcal{A}$.}
    \label{fig:frames}
  \end{center}
\end{figure}

From Eq.~\eqref{eq:14} we can calculate the explicit form of the
Wigner matrix $R_{(\Lambda(\vec{w}), u_A)}$,
\begin{equation}
  \label{eq:37}
  R_{(\Lambda(\vec{w}), u_A)} = B^{-1}(u_B) \Omega(\vec{w}, u_A) B(u_A),
\end{equation}
where
\begin{subequations}
  \label{eq:38}
  \begin{gather}
    \label{eq:38a}
    B(u) = I + \frac{u^0}{1 + u^0} \vec{u} \otimes \vec{u}^T,\\
    \label{eq:38b}
    \Omega(\vec{w}, u) = I 
    + \frac{1}{1 + \sqrt{1 + \vec{w}^2}} \vec{w} \otimes \vec{w}^T 
    - u^0 \vec{w} \otimes \vec{u}^T,
  \end{gather}
\end{subequations}
and by means of Eq.~\eqref{eq:3}
\begin{equation}
  \label{eq:39}
  u^0_B = \frac{u^0_A}{w^0}, \qquad
  \vec{u}_B = \vec{u}_A - \frac{\vec{w}}{u^0_A} 
  \frac{1 + w^0}{1 + \sqrt{1 + \vec{w}^2}}.
\end{equation}
The corresponding velocities of the preferred frame with respect to
frames $\mathcal{A}$ and $\mathcal{B}$ are $\bm{\sigma}_A =
\vec{u}_A/u^0_A$ and $\bm{\sigma}_B = \vec{u}_B/u^0_B$, respectively, while
the velocity of the frame $\mathcal{B}$ with respect to $\mathcal{A}$
is $\vec{V} = \vec{w}/w^0$ (see Fig.~\ref{fig:frames}).
We remark that it is possible to express $R_{(\Lambda(\vec{w}), u_A)}$,
given by Eq.~\eqref{eq:37}, by these velocities as well as by the
corresponding velocities in the Einstein's synchronization with the
help of Eqs.~\eqref{eq:5} because it is only a reparametrization on
the level of \emph{classical} parameters, so it cannot affect the
quantum correlations: They are still dependent on the corresponding
velocities of PF with respect to the observers.  Indeed, as it was
discussed in Sec.~\ref{sec:poinc-covar-quant}, the QMs built up on the
different PFs are not unitary equivalent.  Thus the dependence of
quantum correlation functions on the velocities of PF is unremovable
because it is a pure quantum phenomenon.

Now, the correlation function \eqref{eq:36a} is
\begin{equation}
  \label{eq:40}
  C(\vec{a}, \vec{b}) \propto \vec{a} \cdot B(u_A) \Omega^T(\vec{w}, u_A) B^{-1}(u_B) \vec{b},
\end{equation}
where $B$ and $\Omega$ are given by the formulas~\eqref{eq:38}.  Note, that
the correlation function given by Eq.~\eqref{eq:40} depends on the
choice of PF, i.e., the two correlation functions, say $C(\vec{a},
\vec{b})$ obtained for PF with the four-velocities $u_A$ and $u_B$
with respect to the observers and $\tilde{C}(\vec{a}, \vec{b})$
obtained under \emph{another choice} of PF with the four-velocities
$\tilde{u}_A$ and $\tilde{u}_B$ with respect to the observers,
\emph{do not turn into themselves} when expressing $u_A$ and $u_B$ by
$\tilde{u}_A$ and $\tilde{u}_B$ or vice versa.  This property, related
to the above mentioned nonequivalence of QMs built on different PFs,
can be used to set up the experiments testing the existence and/or
identification of the quantum preferred frame.

Let us discuss some special cases of Eq.~\eqref{eq:40}.
\begin{enumerate}[(1)]
\setlength{\leftmargin}{0pt}
\item $\vec{w} = \vec0$ (i.e.\ $\vec{V} = \vec0$).  In this case both
  measurements are performed in the same inertial frame.  It follows
  from Eq.~\eqref{eq:40}, that the correlation function has the
  standard nonrelativistic form in this case,
  \begin{equation}
    \label{eq:41}
    C(\vec{a}, \vec{b}) \propto \vec{a} \cdot \vec{b} = \cos \theta_{\vec{a}\vec{b}},
  \end{equation}
  as it should be expected.  We would like to point out that the
  correlation function for relativistic EPR particles calculated in
  Refs.~\cite{czachor97a,czachor97b} contains corrections of the order
  $(\text{particle velocity}/c)^2$ to Eq.~\eqref{eq:41}.  It would be
  interesting to verify both the predictions experimentally.
\item $\vec{u}_A = \vec0$ or $\vec{u}_B = \vec0$.  In this case one of
  the observers performs his/her measurement in the preferred frame.
  With the help of Eqs.~\eqref{eq:40} and~\eqref{eq:39} we find that
  \begin{equation}
    \label{eq:42}
    C(\vec{a}, \vec{b}) \propto \vec{a} \cdot \vec{b},
  \end{equation}
  that is, we get the standard nonrelativistic formula.
\item Let us assume that the velocities $\bm{\sigma}_A$ and $\bm{\sigma}_B$ are
  small, i.e.\ $|\bm{\sigma}_A| \ll 1$ and $|\bm{\sigma}_B| \ll 1$.  Such a
  situation occurs if the quantum-mechanical preferred frame coincides
  with the CBRF and the observers' velocities are similar to the
  velocity of the solar system (i.e., $|\bm{\sigma}_A|$ and $|\bm{\sigma}_B| \sim
  10^{-3}$).  In this case
  \begin{displaymath}
    \begin{split}
      R_{(\Lambda(\vec{w}), u_A)} &\simeq I + \frac{\vec{V} \otimes \bm{\sigma}_A^T
        - \bm{\sigma}_A \otimes \vec{V}^T}{2}\\
      &= I + \frac{\bm{\sigma}_A \otimes \bm{\sigma}_B^T - \bm{\sigma}_B \otimes \bm{\sigma}_A^T}{2},
    \end{split}
  \end{displaymath}
  so
  \begin{equation}
    \label{eq:43}
    C(\vec{a}, \vec{b}) \propto \vec{a} \cdot \vec{b} 
    + \frac{(\vec{a} \times \vec{b}) \cdot (\bm{\sigma}_A \times \bm{\sigma}_B)}{2}.
  \end{equation}
  Here $\bm{\sigma}_A$, $\bm{\sigma}_B$ and $\vec{V}$ are related by the
  approximate formula $\bm{\sigma}_B \simeq \bm{\sigma}_A - \vec{V}$.  In the
  formula~\eqref{eq:43} the velocities are given in the absolute
  synchronization scheme but up to the fourth-order corrections they
  have the same form in terms of velocities $\vec{V}_E, {\bm{\sigma}_A}_E,
  {\bm{\sigma}_B}_E$ defined in the Einstein's synchronization scheme (as
  it was mentioned above the reparametrization of the \emph{classical}
  velocities cannot affect the distinguishing of the quantum preferred
  frame, i.e., the quantum correlation function is still dependent on
  the corresponding velocities of PF with respect to the observers).
  The deviation from the standard formula when $\vec{a}$ and $\vec{b}$
  are perpendicular is shown in the Fig.~\ref{fig:corr}.
  \begin{figure}
    \begin{center}
      \includegraphics[width=\columnwidth]{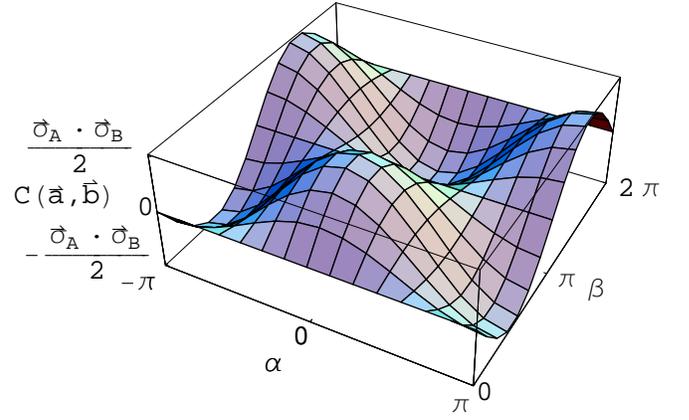}
      \caption{Correlation function $C(\vec{a},\vec{b})$ given by 
        Eq.~\eqref{eq:43} for the case when $\vec{a} \perp \vec{b}$.  Here
        $\alpha$ is the angle between $\vec{a} \times \vec{b}$ and $\bm{\sigma}_A \times
        \bm{\sigma}_B$, and $\beta$ is the angle between $\bm{\sigma}_A$ and $\bm{\sigma}_B$.}
      \label{fig:corr}
    \end{center}
  \end{figure}
  
  Note that it follows from Eq.~\eqref{eq:43} that the corrections to
  the standard formula are of the order 2 in velocities.  With the
  identification of the preferred frame with the CBRF and
  $\mathcal{A}$ and $\mathcal{B}$ with the solar system these
  corrections are of the order $10^{-6}$.  Therefore, we can imagine
  an experiment testing this identification based on the measurement
  of the quantum correlations under the condition that the vectors
  $\vec{a}$ and $\vec{b}$ are perpendicular.  In this case the
  standard part of the correlation function vanishes and only the
  effect caused by the existence of the quantum preferred frame
  remains [see Eq.~\eqref{eq:43} and Fig.~\ref{fig:corr}].  Now,
  unlike in the standard EPR experiments, we should not measure the
  dependence of the correlation function on the angle between the
  vectors $\vec{a}$ and $\vec{b}$, but rather its dependence on the
  change of the velocities of PF, $\bm{\sigma}_A$ and $\bm{\sigma}_B$, caused by
  the movement of the Earth.
\item Finally we consider the case when velocities of the preferred
  frame are high.  Denoting $\vec{u}_A/|\vec{u}_A| = \vec{n}_A$,
  $\vec{u}_B/|\vec{u}_B| = \vec{n}_B$ we obtain in this case
  \begin{multline*}
    R_{(\Lambda(\vec{w}),u_A)} \simeq I - \left(\vec{n}_A \otimes \vec{n}_A^T
      + \vec{n}_B \otimes \vec{n}_B^T + \vec{n}_A \otimes \vec{n}_B^T \right.\\
    \left.- (1 + 2 \vec{n}_A \cdot \vec{n}_B) \vec{n}_B \otimes
      \vec{n}_A^T\right) \left(1 + \vec{n}_A \cdot \vec{n}_B\right)^{-1},
  \end{multline*}
  hence,
  \begin{multline}
    \label{eq:44}
    C(\vec{a}, \vec{b}) \propto \vec{a} \cdot \vec{b} - \frac{1}{1 + \vec{n}_A \cdot
      \vec{n}_B} \left[(\vec{a} \cdot \vec{n}_A) (\vec{b} \cdot \vec{n}_A)
      + (\vec{a} \cdot \vec{n}_B) (\vec{b} \cdot \vec{n}_B)\right.\\
    \left.+ (\vec{a} \cdot \vec{n}_B) (\vec{b} \cdot \vec{n}_A) - (1 + 2
      \vec{n}_A \cdot \vec{n}_B) (\vec{a} \cdot \vec{n}_A) (\vec{b} \cdot
      \vec{n}_B)\right].
  \end{multline}
\end{enumerate}

We point out that the simultaneity of the measurements ($t_A = t_B$)
is defined in the \emph{corresponding absolute synchronization scheme}
related to the choice of the PF \footnote{If the two measurements are
  performed simultaneously (in absolute synchronization) in the places
  at the distance $l$, the difference of their time in Einstein's
  synchronization is $|\Delta\tau| \leq \sigma_E l/c^2$ [cf.\ Eqs.~\eqref{eq:4}],
  i.e., if PF is CBRF then for $l \sim 1\,\mathrm{km}$ we have $\Delta\tau \lesssim
  1\,\mathrm{ns}$.}.

\section{Conclusions}
\label{sec:conclusions}

In the framework of the Lorentz-covariant quantum mechanics with the
preferred frame one can build the formalism that allows to calculate
correlation function in the EPR-type experiments [see
Eqs.~\eqref{eq:29} and \eqref{eq:30}] performed in moving inertial
frames.  We would like to point out that our results are the exact EPR
correlation functions obtained for Lorentz-covariant quantum
mechanical systems in moving frames under physically acceptable
conditions, i.e., taking into account the localization of the
particles during the detection and using the spin operator with proper
transformation properties under the action of the Lorentz group.

We applied the general result to the case of simultaneous measurements
of the spin component for bipartite spin-1/2 system done by the
spatially bounded detectors.  The resulting correlation function is
proportional to $\vec{a} \cdot R^T_{(\Lambda,u_A)} \vec{b}$, where $\vec{a}$ and
$\vec{b}$ are the direction vectors and $R_{(\Lambda,u_A)}$ is the Wigner
rotation matrix associated with the Lorentz transformation $\Lambda$
connecting the frames of the detectors.  Next we have studied the
limiting cases of this particular correlation function and have shown
that in the case when both measurements are performed in the same
inertial frame we obtain the standard nonrelativistic result that the
correlation function is proportional to the scalar product of the
direction vectors.  This result also holds if one of the measurements
is performed in the preferred frame.  We have also found the limit of
the correlation function for small velocities and shown that it leads
to the correction of the second order in velocities to the standard
$\vec{a} \cdot \vec{b}$ relation.  On the other hand, the correlation
function for the very high velocities of the PF with respect to the
observers depends only on the directions of movement of the PF.

It is important to stress that the exact EPR correlation
function~\eqref{eq:29} depends on the PF velocity in an essential way,
i.e., this dependence cannot be removed by expressing the correlation
function by classical quantities (velocities) given in the Einstein's
synchronization scheme.  This means that \emph{the Lorentz-covariant
  quantum mechanics must distinguish a preferred frame}.  The above
results can be used to propose a realistic experiment which can answer
the question of the existence of quantum-mechanical preferred frame
(and its possible identification with the CBRF).  A more exhaustive
discussion of this problem as well as an analysis of the subtle
question concerning the synchronization of clocks in the experimental
setup will be given in the forthcoming paper.

\begin{acknowledgments}
  One of the authors (JR) is grateful to Marek Czachor, Ryszard
  Horodecki, Valerio Scarani, and Anton Zeilinger for discussions
  during the second EFS QIT conference in Gda{\'n}sk as well as Harvey R.
  Brown for discussing the conventionality of the synchronization in
  the special relativity.  This work was supported by Lodz University
  Grant No.~267.
\end{acknowledgments}


\end{document}